\documentclass[aps,prl,twocolumn,showpacs]{revtex4}
\usepackage{stmaryrd}
\usepackage{amsfonts}
\usepackage{mathrsfs}
\usepackage{amsmath}
\usepackage{amscd}
\usepackage{graphicx}
\usepackage{booktabs}
\usepackage{leftidx}
\begin{document}
\title{Multi-photon quantum communication in quantum networks}
\author{Wei Qin$^{1,2,3}$}
\author{Chuan Wang$^4$}
\author{Ye Cao$^{1,2,3}$}
\author{Gui Lu Long$^{1,2,3}$}\protect\thanks{Corresponding author£º gllong@mail.tsinghua.edu.cn}
\address{$^1$State Key Laboratory of Low-Dimensional Quantum Physics and Department of Physics, Tsinghua University, Beijing 100084,
China\\
$^2$Collaborative Innovation Center of Quantum Matter, Beijing 100084, China\\
$^3$Tsinghua National Laboratory for Information Science and Technology, Tsinghua University, Beijing 100084, China\\
$^4$School of Science, Beijing University of Posts and Telecommunications, Beijing 100876, China}
\begin{abstract}
We propose and analyze a multi-photon state coherent transport protocol in a coupled-resonator quantum network. A multi-photon SWAP gate between two antipodes can be achieved  with neither external modulation nor coupling strength engineering. Moreover, we extend this result to a coupled-resonator chain of arbitrary length with different coupling strengths. Effects of decoherence via quantum non-demolition interaction are studied with sources including vacuum quantum fluctuation and bath thermal excitations when the bath is in the thermal equilibrium state. These observations are helpful to understand the decoherence effects on quantum communication in quantum coupled-resonator systems.
\end{abstract}
\maketitle


Coherent transport of quantum information between two remote qubits is of central importance in quantum information processing (QIP). There have been many studies of connecting remote solid qubits and realizing transport in various systems, including flux qubits in superconductors \cite{superconductor1,superconductor2}, phonons in ion traps \cite{ion1,ion2} and nuclear spins in nuclear magnetic resonance (NMR) \cite{NMR}. Because of negligible interaction between separated photons, high-speed transmission with low dissipation in optical fibers and compatibility with class telecommunication fiber technology, an optical quantum network has become one of the most promising candidates for scalable QIP in the past decades \cite{NTW}. In this case, photon coherent transport is of both fundamental and practical importance to perform communication between two nodes \cite{Tras}. Knill {\it et al.} showed that an efficient quantum computation can be implemented with single-photon sources, single-photon detectors and linear optics alone \cite{knill}, however, the complexity of required networks is daunting. Furthermore, the narrow spectral bandwidth in the conventional single-photon cavities with high $Q$ factor will decrease the single-photon detection efficiency \cite{lund}. For these reasons, most of alterative and feasible approaches have been proposed by encoding quantum information on multi-photon fields to overcome these limitations \cite{CHS1,CHS2,CHS3,CHS4,CHS5,CHS6, CHS7,CHS8,CHS9,CHS10}.

Prior work on quantum communication in quantum networks has commonly focused on either spin or single-photon qubits \cite{CQST,Chu,PJP}, while the study of coupled-resonator chains with continuous-variable quantum states has attracted much attention \cite{CV1,CV2,CV3,CV4,CV5}, and the mapping of quantum states between photons and atoms has been implemented in experimental systems \cite{map1,map2,map3,map4}, as a critical requirement for distributed quantum information. In this Letter we investigate the multi-photon state coherent transport between two antipodes in coupled-resonator quantum networks based on Cartesian products of graph theory \cite{CQST,GTH}. A chain of two or three resonators can work as basic building blocks to build quantum coupled-resonator networks which are multiple Cartesian products of either of the two simple chains. This can achieve a perfect multi-photon SWAP gate after a period of time evolution
on a hypercubic structure that is one of those commonly used in networks and a direct generalization of spin chains. The method swaps the arbitrary bosonic states of two antipodes resonators under time evolution, which is determined by the natural dynamics and requires neither external modulation of Hamiltonian nor inter-resonator coupling strength engineering. Its essence is that a perfect SWAP operation is allowable for a chain of either two or three resonators. As a consequence, we extend this result to a coupled-resonator chain of arbitrary length and a mirror inversion of bosonic states with respect to its center is implemented. Also the optimal time over these coupled-resonators is independent of the distance between two remote nodes and the speedup of the perfect state transfer is possible.

The interaction between a realistic quantum system and its surrounding environment is hardly avoidable. The proposed protocol with perfect quantum state transfer occurs in a closed system or an ideal condition without decoherence. Thus it is necessary to study the decoherence effects on such protocol. The decoherence is characterized by a pure dephasing model for an open system coupled to a bosonic bath via quantum non-demolition interaction \cite{DP1}. We assume that the decoherence effects on each eigenmode of the network are identical and the decoherence occurs between the occupancy number bases in Fock space. After unitary evolution and in Heisenberg picture, a SWAP gate under decoherence is achieved with two additional phase factors. In a special case where the bath is in the thermal equilibrium state, the sources of decoherence effects on the gate include vacuum quantum fluctuation and bath thermal excitations at finite temperature. Observing these will help us to understand the decoherence effects on quantum communication in quantum coupled-resonator systems.

\begin{figure}[!ht]
\begin{center}
\includegraphics[width=8.5cm,angle=0]{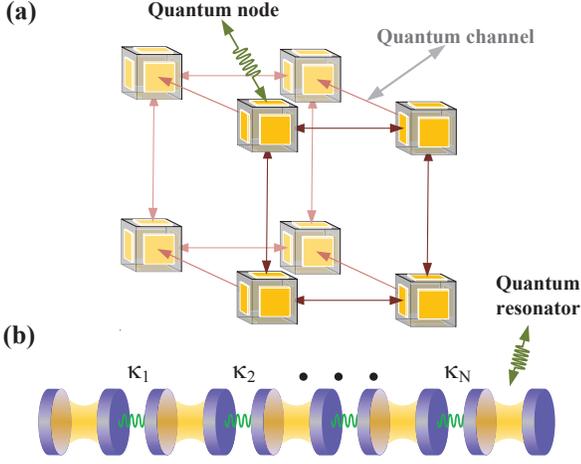}
\caption{(a) A cubic quantum network is the $3$-fold Cartesian product of two-resonator chain with uniform coupling strength. A multi-photon SWAP gate between the two nodes along each main diagonal is achieved at the optimal time. (b) A coupled-resonator chain with different coupling strengths. Its coupling strength distribution is characterized by the $x$ component of an angular momentum operator to implement a mirror inversion of arbitrary bosonic states with respect to the center of the chain.}\label{f1}
\end{center}
\end{figure}

The Hamiltonian of a coupled-resonator quantum network described by a graph $\mathcal{G}$ is
\begin{equation}\label{Hs}
H_{S}=\sum_{u=1}^{N} \Omega a^{\dag}_{u}a_{u}+\sum^{N}_{u,v=1}K_{uv}(\mathcal{G})a^{\dag}_{u}a_{v},
\end{equation}
where $a_{u}$ and $a_{u}^{\dag}$ are the bosonic annihilation and creation operators for the single mode
resonators with frequency $\Omega$ at node $u$, $K_{uv}(\mathcal{G})=\kappa A_{uv}(\mathcal{G})$ represents the coupling strength between nodes $u$ and $v$. $A(\mathcal{G})$ is the adjacency matrix of graph $\mathcal{G}$,
$A_{uv}(\mathcal{G})=1$ if the two nodes $u$ and $v$ are adjacent; otherwise $A_{uv}(\mathcal{G})=0$. It means that only the nearest neighbor (NN) coupling is considered. Since $A$ is a symmetric matrix, its diagonalization occurs through
an orthogonal transformation $U$ as $D=UAU^{\dag}$ with $D_{kl}=\lambda_{k}\delta_{kl}$.
This transformation yields $H_{S}=\sum^{N}_{k=1}\varepsilon_{k}f_{k}^{\dag}f_{k}$,
where $f_{k}=\sum_{u=1}^{N}U_{ku}a_{u}$ is the Bogoliubov transformation and $\varepsilon_{k}=\Omega+\kappa\lambda_{k}$. The bath Hamiltonian consisting of harmonic oscillators with infinite modes is $H_{B}=\sum_{j}\omega_{j}b^{\dag}_{j}b_{j}$, where $b_{j}$ ($b^{\dag}_{j}$) are the bosonic annihilation (creation) operators for the modes of frequencies $\omega_{j}$ and $j=1,2,\cdots,\infty$. The interaction between the quantum network and the bath is characterized by a simple decoherence model with non-demolition Hamiltionian $H_{I}=R \otimes X$ \cite{DP1} , where $R=\sum_{k=1}^{N}r_{k}f_{k}^{\dag}f_{k}$ depends on the network variables and the quantum noise operator $X=\sum_{j}\xi_{j}b_{j}+H.c.$ on the bath variables. The dynamics of composite system is driven by the total Hamiltonian $H=H_{S}+H_{B}+H_{I}$. Since $[H_{I},H]=0$ and $[H_{B},H]\neq0$, the energy exchange is unallowable between the network and its surrounding bath, and an irreversible process of information loss happens.

In Heisenberg picture, the Heisenberg equation of motion, $\dot{O}(t)=i[H,O(t)]$, governs the time evolution of an operator. By applying this equation, the exact solutions of operators $f^{\dag}_{k}$ and $b_{j}^{\dag}$ can be found \cite{DP2}
\begin{equation}\label{ft}
f_{k}^{\dag}(t)=e^{i\varepsilon_{k}t}e^{ir_{k}[Z(t)-F(t)R_{0}]}f_{k}^{\dag}(0)
\end{equation}
and
\begin{equation}
b_{j}^{\dag}(t)=e^{i\omega_{j}t}b_{j}^{\dag}(0)+i\xi_{j}\eta^{*}_{j}(t)R_{0},
\end{equation}
where $Z(t)=\sum_{j}[\xi_{j}\eta_{j}(t)b_{j}(0)+H.c.]$ is the phase operator with $\eta_{j}(t)=i(e^{-i\omega_{j}t}-1)/\omega_{j}$, $F(t)=2\int d\omega J(\omega)(t-sin\omega t/\omega)/\omega$ is a c-number with the bath spectral density function $J(\omega)=\sum_{j}|\xi_{j}|^{2}\delta(\omega-\omega_{j})$, and $R(t)=R_{0}$ as a result of its conservation, i.e., $[H,R(t)]=0$. $r_{k}$ is a parameter to measure the decoherence effects on the $k$th mode of network. We assume that the effects on each mode are identical, $r_{k}=r$, and the decoherence occurs in Fock space. After the inverse Bogoliubov transformation and time evolution, the creation operator $a^{\dag}_{m}$ becomes
$a_{m}^{\dag}(t)=e^{iY(t)}\widetilde{a}_{m}^{\dag}(t)$. The term $\widetilde{a}_{m}^{\dag}(t)=\sum_{k}U_{km}e^{i\varepsilon_{k}t}f^{\dag}_{k}(0)$ is the network free evolution and $Y(t)=r[Z(t)-F(t)R_{0}]$ represents the decoherence effects. Upon introducing Bogoliubov transformation again, $\widetilde{a}_{m}^{\dag}(t)$ is transformed to $\widetilde{a}_{m}^{\dag}(t)=e^{i\Omega t}\sum^{N}_{u=1}(e^{i\kappa A t})_{um}a_{u}^{\dag}$. Thus the time evolution of quantum states is driven by the adjacency matrix of the network, which is analogous to the spin networks \cite{CQST}.

A chain of two or three resonators can act as basic building blocks to build coupled-resonator networks which are multiple Cartesian product of either of the two simple chains \cite{CQST}. The two-resonator chain is denoted by $\mathcal{G}_{1}$ and the three-resonator one by $\mathcal{G}_{2}$. The three-fold Cartesian product of $\mathcal{G}_{1}$ is a cubic network as shown in Fig. \ref{f1}(a). The relations of adjacency
matrices $A(\mathcal{G})$ and $A(\mathcal{G}_{\theta})$ after $g$-fold Cartesian product obey the rules of Kronecker product $A(\mathcal{G})=\sum_{j=0}^{g-1}I^{\otimes j}\otimes A(\mathcal{G}_{\theta})\otimes
I^{\otimes (g-j-1)}$ with an identity matrix $I$ and $\theta=1, 2$. Consequently, $e^{A(\mathcal{G})}=[e^{A(\mathcal{G}_{\theta})}]^{\otimes g}$, and $A(\mathcal{G_{\theta})}$ determines the evolution of network $\mathcal{G}$. After evolution and at the optimal time $t=\tau_{\theta}\equiv \pi/2^{1/\theta}\kappa$, $[e^{i\kappa A(\mathcal{G})\tau_{\theta}}]_{um}=i^{\theta g}\delta_{u,N+1-m}$ gives that
\begin{equation}
a_{m}^{\dag}(\tau_{\theta})=P_{0}P_{1}a^{\dag}_{N+1-m},
\end{equation}
demonstrating the relations between the state of node $m$ at $t=0$ and that of node $N+1-m$ at $t=\tau_{\theta}$. Here, the phase $P_{0}=e^{i\Omega\tau_{\theta}}i^{\theta g}$ arises from the free evolution of network and $P_{1}=e^{iY(\tau_{\theta})}$ from the decoherence. Actually, it is a perfect SWAP gate between the two antipodes in the absence of decoherence. It requires neither external manipulation nor coupling strength engineering. For simplicity, we take the cubic quantum network as an example, as shown in Fig. \ref{f1}(a), the time evolution swaps arbitrary bosonic states of the two antipodes along each main diagonal at the optimal time.

In ideal conditions without decoherence, the proposed method can potentially allow for the realization of a perfect SWAP gate with coherent-state qubits between two antipodes in the quantum network. We consider an initial state $|\Phi\rangle_{ini}=\prod^{N}_{u=1}|\alpha_{u}\rangle_{u}$, where $|\alpha_{u}\rangle_{u}$ is the coherent state with amplitude $\alpha_{u}$ at node $u$. The coherent state can be produced by a displacement operator $D(\alpha)=e^{-|\alpha|^{2}/2}e^{\alpha a^{\dag}}e^{-\alpha^{*}a}$ displacing the vacuum state $|0\rangle$, $|\alpha\rangle=D(\alpha)|0\rangle$. After evolution in Heisenberg picture and inversion back to Schr\"{o}inger picture, the final quantum network state becomes
\begin{equation}
|\Phi\rangle_{fin}=\prod^{N}_{u=1}|\beta_{u}\rangle_{N+1-u}
\end{equation}
with $\beta_{u}=P_{0}^{*}\alpha_{u}$. The coherent-state SWAP gate between nodes $m$ and $N+1-m$ is achieved. In addition to the coherent state, the coherent transport of the multi-photon entangled states can be implemented by means of our scheme with an alterative Hamiltonian
\begin{equation}\label{pH}
H_{a}=\sum_{\sigma}\left[\sum_{u=1}^{N} \omega a^{\dag}_{u,\sigma}a_{u,\sigma}+\sum^{N}_{u,v=1}K_{uv}(\mathcal{G})a^{\dag}_{u,\sigma}a_{v,\sigma}\right],
\end{equation}
where $\sigma=h,v$ are the photon polarization states with $h$ represented by horizontal polarization and $v$ by vertical.
The creation operator with polarization state $\sigma$ is likewise transformed to $\widetilde{a}_{m,\sigma}^{\dag}(\tau_{\theta})=P_{0}a_{N+1-m,\sigma}^{\dag}$ at the optimal time $\tau_{\theta}$. Without loss of generality and for simplicity, an initial state $|\Phi'\rangle_{ini}=\prod^{N}_{u=1}|\Psi'\rangle_{u}$ with
\begin{equation}
|\Psi'\rangle_{u}=\left(|h\rangle^{\otimes M_{u}}+|v\rangle^{\otimes M_{u}}\right)_{u}/\sqrt{2}
\end{equation}
is taken as an example. The time evolution under the Hamiltonian $H_{a}$ swaps the multi-photon entangled states of two antipodes with the finial state
\begin{equation}
|\Phi'\rangle_{fin}=P_{E}\left(\prod^{[N/2]}_{u=1}\text{SWAP}_{u,N+1-u}\right)|\Phi'\rangle_{ini},
\end{equation}
where $P_{E}=\prod^{N}_{u=1}(P_{0}^{*})^{M_{u}}$ is an additional phase.

Besides the case of single mode resonators, the implementation of multi-mode resonators of frequencies $\Omega_{\vartheta}$ is directly analogous. In multi-mode quantum networks, the Hamiltonian is $H_{S}=\sum_{\vartheta}H_{\vartheta}$ with $H_{\vartheta}=\sum_{u=1}^{N} \Omega_{\vartheta} a^{\dag}_{u,\vartheta}a_{u,\vartheta}+\sum^{N}_{u,v=1}K_{uv}(\mathcal{G})a^{\dag}_{u,\vartheta}a_{v,\vartheta}$. Since $[H_{\vartheta},H_{\vartheta'}]=\delta_{\vartheta\vartheta'}$, it is possible to have $\widetilde{a}_{m,\vartheta}^{\dag}(\tau_{\theta})=P_{0,\vartheta}a^{\dag}_{M+1-m,\vartheta}$ with $P_{0,\vartheta}=e^{i\Omega_{\vartheta}\tau_{\theta}}i^{\theta g}$.

{\it Extensions.}\text{---}While the case of a multi-dimensional hyercube has been chosen to focus on, we extend this result to a one-dimensional (1D) coupled-resonators and the realization of such resonator chain can be explored in many physical systems  \cite{OMC,CA1,CA2,SC,NS,DPC}. If $K_{uv}(\mathcal{G})=\kappa_{u-1}\delta_{u,v+1}+\kappa_{u}\delta_{u,v-1}$ in the Hamiltonian of Eq. (\ref{Hs}), the Hamiltonian describes a 1D coupled-resonator system with different coupling strengths as shown in Fig. \ref{f1}(b). The coupling strength distribution matrix $K$ is identical to the representation of a Hamiltonian $H'=\lambda J_{x}$ through pre-engineering the inter-resonator coupling strengths as $\kappa_{u}=\lambda\sqrt{u(N-u)}/2$, where $J_{x}$ is the $x$ component of a fictitious angular momentum operator $J=(N-1)/2$ and $\lambda$ is some constant \cite{CQST}.

The isomorphism of $su(2)$ and $so(3)$ Lie algebras gives the remarkable result that $SU(2)$ and $SO(3)$ Lie groups are locally isomorphic, and the commutation relations of an arbitrary angular momentum operator can be reduced to those of harmonic oscillator operators in Schwinger picture, e.g., $J_{x}$ can be rewritten in terms of two bosonic operators as $J_{x}=(c^{\dag}_{1}c_{2}+c_{1}c^{\dag}_{2})/2$ \cite{JSch}. $H'$ can be viewed as a Hamiltonian for the two resonators with coupling strength $\lambda/2$ and perfect quantum state transfer between the two resonators is possible, which gives that $e^{iH'\tau'}c^{\dag}_{1}e^{-iH'\tau'}=ic^{\dag}_{2}$ and $e^{iH'\tau'}c^{\dag}_{2}e^{-iH'\tau'}=ic^{\dag}_{1}$ at the optimal time $t=\tau'\equiv\pi/\lambda$. It yields $\langle m_{z}'|e^{i \lambda J_{x} \tau'}|m_{z}\rangle=i^{2J}\delta_{m'_{z}m_{z}}$ and $(e^{iK\tau'})_{um}=i^{N-1}\delta_{u,N+1-m}$, where $m_{z},m'_{z}=-J,-J+1,...,J-1,J$ and $|m_{z}\rangle=(a^{\dag}_{1})^{J+m_{z}}(a^{\dag}_{2})^{J-m_{z}}/\sqrt{(J+m_{z})(J-m_{z})}|0\rangle$. Thus $\widetilde{a}_{m}^{\dag}(t)$ becomes
\begin{equation}
\widetilde{a}_{m}^{\dag}(\tau')=P'_{0}a^{\dag}_{N+1-m},
\end{equation}
where $P'_{0}=e^{i\Omega \tau'}i^{N-1}$. As desired, the SWAP gate between sites $m$ and $N+1-m$ is achieved under time evolution. The noise effect is $P'_{1}=e^{iY(\tau')}$ when the decoherence is present.

{\it Decoherence in thermal equilibrium state.}\text{---}When the unavoidable bath is in the thermal equilibrium state, its variables are distributed in an uncorrelated thermal equilibrium mixture of states and the density matrix satisfies Boltzmann distribution $\rho_{B}=e^{-H_{B}/T}/Z$, where $T$ represents the temperature and $Z=tr(e^{-H_{B}/T})$ is the partial function. A density operator $\rho$ can be expressed in terms of coherent states in coherent-state representation $\rho=\prod_{j}\int \rho_{j}(\alpha_{j},\alpha_{j}^{*})|\alpha_{j}\rangle\langle \alpha_{j}|d^{2}\alpha_{j}$, and
\begin{equation}
\rho_{j}(\alpha_{j},\alpha_{j}^{*})=tr[\rho\delta(\alpha_{j}^{*}-a^{\dag}_{j})\delta(\alpha_{j}-a_{j})]
\end{equation}
builds a connection between the classical and quantum coherence theory \cite{Qoptic}. For the thermal equilibrium state, $\rho_{j}(\alpha_{j},\alpha_{j}^{*})=e^{-|\alpha_{j}|^{2}/\langle n_{j}\rangle}/\pi\langle n_{j} \rangle$ with the average excitation number $\langle n_{j} \rangle=(e^{\omega_{j}/T}-1)^{-1}$ in the modes of frequencies $\omega_{j}$. The initial state of composite system is a direct product $\rho(0)=\rho_{S}(0)\otimes\rho_{B}$, and the density matrix is $\rho_{S}(0)=|\psi(0)\rangle\langle\psi(0)|$ with a generic state $|\psi(0)\rangle=\prod^{N}_{u}\sum_{n_{u}}c_{n_{u}}|n_{u}\rangle_{u}$.

After evolution and tracing out the variables of the bath, we have the reduced density matrix of quantum network at the optimal time $\rho_{S}(\tau')=\sum_{\mathbf{n},\mathbf{n'}} D_{n,n'}(T)\rho_{\mathbf{n},\mathbf{n'}}(\tau')$ with two vectors $\mathbf{n}=\mathbf{n}(n_{1},...,n_{N})$ and $\mathbf{n'}=\mathbf{n'}(n'_{1},...,n'_{N})$
in a chain of resonators, it is directly analogous in the case of hypercubic networks.
$D_{n,n'}(T)$ is an expected value of a displacement operator $D_{j}(\beta_{jn'n})$ in the thermal equilibrium state
\begin{equation}\label{DTnm}
D_{n,n'}(T)
=\prod_{j}\int d^{2}\alpha_{j}\rho_{j}(\alpha_{j},\alpha_{j}^{*})
\langle\alpha_{j}|D_{j}(\beta_{jn'n})|\alpha_{j}\rangle,
\end{equation}
where $\beta_{jn'n}=-i(n-n')r\xi_{j}^{*}\eta_{j}(\tau')$,  $D_{j}(\beta_{jn'n})=e^{\beta_{jn'n}b^{\dag}_{j}-\beta^{*}_{jn'n}b_{j}}$, $n=\sum^{N}_{u=1}n_{u}$ and $n'=\sum^{N}_{u=1}n'_{u}$. $D_{n,n'}(T)=D_{n,n'}^{[0]}D_{n,n'}^{[T]}$ includs the vacuum quantum fluctuation $D_{n,n'}^{[0]}=\Pi_{j}e^{-z_{nn';j}(\tau')/2}$ and the bath thermal excitations $D_{n,n'}^{[T]}=\prod_{j}e^{-g_{j}(T)}$ with $g_{j}(T)=
z_{nn';j}(\tau')/(e^{\omega_{j}/T}-1)$ and $z_{nn';j}(\tau')=|\beta_{jn'n}|^{2}$. $\rho_{\mathbf{n},\mathbf{n'}}(\tau')$ is
\begin{equation}
\rho_{\mathbf{n},\mathbf{n'}}(\tau')=\prod^{N}_{u=1}c_{n_{u}}(\tau')c^{*}_{n'_{u}}(\tau')|n_{u}\rangle\langle n'_{u}|_{N+1-u},
\end{equation}
where $c_{n_{u}}(\tau')=(P'^{*}_{0})^{n_{u}}e^{-in_{u}(n_{u}+1)r^{2}F(-\tau')/2}c_{n_{u}}$. As an illustration, the Ohmic spectral
density of bath is taken by $J(\omega)=\gamma\omega e^{-\omega/\Gamma}$ with a dimensionless coupling constant $\gamma$ and the bath's response frequency $\Gamma$ \cite{OHS1,OHS2}. We have that
$D^{[0]}_{n,n'}=(1+\Gamma^{2}\tau'^{2})^{-(n-n')^{2}r^{2}\gamma/2}$
and $\sum_{j}g_{j}(T)=4\gamma(n-n')^{2}g^{2}I(\tau')$ with
\begin{equation}
I(\tau')=\int d\omega\frac{1}{\omega e^{\omega/\Gamma}(e^{\omega/T}-1)}\sin^{2}\frac{\omega \tau'}{2}.
\end{equation}

\begin{figure}[!ht]
\begin{center}
\includegraphics[width=9.0cm,angle=0]{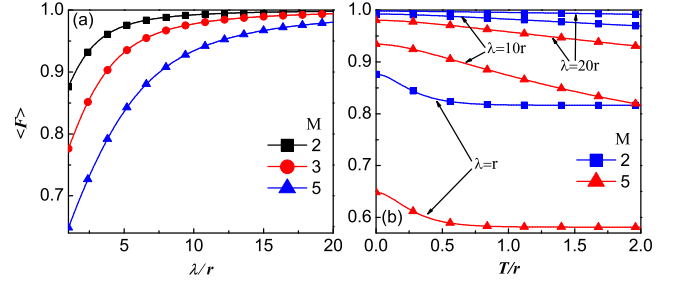}
\caption{The decoherence effects on the proposed multi-photon swap gate in a chain of resonators. $\gamma=1$, $\Gamma=1$. (a) The vacuum quantum fluctuation effects on the swap gate fidelity. (b) The swap gate fidelity as a function of temperature at several coupling parameters. $\langle F\rangle$ is average fidelity, $\lambda$
is a coupling parameter in the coupled-resonator chain, $r$ is a coupling parameter between the
network and the bath, $\gamma$ is a dimensionless coupling constant and $\Gamma$ is the
bath's response frequency in Ohmic spectral density, $T$ is temperature, and $M$
is the dimension of initial state at node $1$ in this special case.}\label{f2}
\end{center}
\end{figure}

A special case is that $|\psi(0)\rangle=(\sum_{n_{1}=0}^{M-1}c_{n_{1}}|n_{1}\rangle_{1})\otimes|\mathbf{0}\rangle$, which means that all of nodes are initialized to vacuum state except for node $1$. The quality of SWAP gate between nodes $1$ and $N$ can be defined by an average fidelity $\langle F\rangle=\sqrt{\overline{\langle\phi_{0}|U^{\dag}\rho(t)U|\phi_{0}\rangle}}$, where the overline indicates average overall possible input state $|\phi_{0}\rangle$ and $U$ is an ideal SWAP gate. It turns out to be at the optimal time
\begin{equation}
\langle F\rangle=\sqrt{\overline{\sum_{n_{1},n'_{1}=0}^{M-1}|c_{n_{1}}|^{2}|c_{n'_{1}}|^{2}D_{n_{1},n'_{1}}(T)}}.
\end{equation}
Figure \ref{f2} shows the fidelity as a function of coupling parameter $\lambda$ and finite temperature $T$ in a chain of coupled-resonators. It is seen that the fidelity decreases with temperature and increases with coupling strength. The strong coupling can partly counteract the decoherence effects to ensure the high fidelity of SWAP gate.

In summary, we have investigated a multi-photon coherent transport protocol in a coupled-resonator quantum network and proposed a perfect multi-photon SWAP gate between two antipodes. The quantum network is the multi-fold Cartesian product of a chain of either two or three resonators and the method requires neither external modulation nor inter-resonator coupling strength engineering. As an extension, we have shown that a multi-photon SWAP gate can be achieved perfectly over a chain of arbitrary length as long as one can pre-engineer inter-resonator coupling strengths. The optimal time is independent of the distance between the two remote parties, and the speedup of state transfer is possible. The decoherence effects on the SWAP operation have been demonstrated explicitly. The sources include the vacuum quantum fluctuation and the bath thermal excitations when the bath is in the thermal equilibrium state. Such observations can help us to deepen our understanding of the decoherence effects on quantum communication in quantum coupled-resonator systems and evaluate the proposed protocol when it works in thermal environment. Additionally, our method can provide the implementation of coherent-state SWAP gate and arbitrary dimensional quantum state transfer based on photons.

We gratefully acknowledge conversations with J. Pearson.
This work was supported by the National Natural Science Foundation of China (GrantNos.
11175094, 91221205,61205117,), the National Basic Research Program of China (2011CB921602), and the Specialized Research Fund for the Doctoral Program of Education
Ministry of China.


\end{document}